\documentclass[a4paper,aps,prl,twocolumn,groupedaddress,showpacs]{revtex4}
\newcommand{\bi}[1]{\mbox{\boldmath{$#1$}}}
\usepackage{graphicx}
\usepackage{epsf}
\begin{document}

\title{Rigorous definition of oxidation states of ions in solids}

\author{Lai Jiang}
\altaffiliation{These two authors contributed equally to this work.}
\author{Sergey V. Levchenko}
\altaffiliation{These two authors contributed equally to this work.}
\author{Andrew M. Rappe}
\email[]{rappe@sas.upenn.edu}
\affiliation{The Makineni Theoretical Laboratories, Department of Chemistry, University
of Pennsylvania, Philadelphia, PA 19104--6323}
\date{\today}

\begin{abstract}
We present justification and rigorous procedure for electron partitioning among
atoms in extended systems. The method is based on wavefunction topology and the modern 
theory of polarization, rather than charge
density partitioning or wavefunction projection, and, as such, re-formulates the concept of oxidation state 
without assuming real-space charge transfer between atoms. This formulation provides 
rigorous electrostatics of finite extent solids, including films and nanowires.

\end{abstract}


\maketitle

The concept of oxidation state (OS) is widely used to predict chemical and spectroscopic properties of compounds, based solely
on the atomic identities and the topology of their bonding~\cite{Jorgensen69}. For instance, the electrostatics of solids is usually
 described by a simple ionic model replacing atoms with point charges equal to the OS.
OS therefore plays an important role in ionic crystals whose
properties are greatly influenced by electrostatics, due to the close packing of ions and the slow decay of the Coulomb interaction with distance.

Real space electron density partitioning among atoms is a traditional way of obtaining OS. It can also be  
used to approximate
 electrostatics and dispersion~\cite{Tkachenko09p073005} in molecules and extended systems.
 However, with the development of wavefunction-based quantum mechanics, it has become widely 
 accepted that there is no rigorous justification for such a partitioning, due to the continuous electronic
distribution. In fact, it has been demonstrated recently that
the assumption about physical transfer of charge upon changing OS is in some cases incorrect due to a ``negative feedback'' mechanism~\cite{Resta08p735,Raebiger08p763}.

Another popular method of assigning OS is the projection of wavefunctions to localized atomic orbitals~\cite{Lowdin50p365,Mulliken55p1833}.
which removes the dependence on charge density. However, it suffers from dependence on basis set and generally produces non-integer OS. 
A projection-based approach provides a way to 
round fractional occupation into integer  OS in metal-ligand systems and
 avoids the ``negative feedback'', but work when strong metal-metal bond is present or the ligand (electron donor) OS is  desired~\cite{Sit11p10259}.

From the examples above, it seems that there is no universal method to assign integer charges to atoms deterministically based on
atomic configuration and electronic structure. Yet in another context, namely electrolysis, ionic charge appears
exactly in quanta. It has also been shown theoretically that the current associated with an atom moving in a torus-like insulator loop
is due to motion of 
integer charges~\cite{Pendry84p1269}. This sheds light on the the idea that OS, being a ground state property, is measurable in a
 process that involves moving the atom in interest.
 
The question still remains of the appropriate quantity to be measured or calculated in order to evaluate  OS. While both charge
density and projected occupation fail to play this role, the modern Berry phase description of polarization 
(or equivalently, localized Wannier functions) has been used to model interface charge~\cite{Bristowe11p081001}, surface stoichiometry~\cite{Levchenko08p256101} and other properties greatly influenced by bulk electrostatics~\cite{Noguera00pR367},  making it a good
candidate for further study.

In this paper, we employ the ideas of
quantized charge transport and modern theory of polarization to develop a rigorous methodology for distributing
electrons among ions in the solid. This scheme  is based solely on 
topology of electronic states rather than electron density.
It also establishes a connection between the concepts of oxidation
state and charge quantization.

For any periodic solid, the
polarization change $\Delta \vec{P}$ 
along an arbitrary path in a parameter space (e.g., in the
space of nuclear coordinates in the adiabatic approximation) can be computed
modulo $e\vec{R}/V$ (where $\vec{R}$ is a lattice vector, and $V$ is 
the volume of the unit cell) from knowledge of the system at initial and 
final points, provided the system remains insulating 
at every point of the path~\cite{Kingsmith93p1651,Vanderbilt93p4442}. 
Furthermore, the uncertainty can be removed by considering 
smaller intervals along the path. 
Here, we focus our attention on a
special subset of such paths, namely, the displacement of an atomic sublattice
by a lattice vector $\vec{R}$. Since the Hamiltonian returns to itself, the
polarization can change only by
\begin{equation}
\Delta \vec{P} = \frac{e}{V}\sum_{i=1}^3 n_i\vec{R}_i
\label{eq:DELTAP}
\end{equation}
where $n_i$ are integers, and $\vec{R}_i$ are the 
lattice vectors defining the unit cell. 

(i) {\em Under certain conditions, $\Delta \vec{P}$ does not depend on the details of the path,
as long as the system stays insulating at every point on the path.}
We note that it is now well established
that the derivatives of polarization with respect to nuclear positions,
\begin{equation}
\left. Z^*_{i,\alpha \beta}=V \frac{\partial P_{\alpha}}{\partial \Re _{i\beta}}\right|_{{\cal E}=0}
\label{eq:ZSTAR}
\end{equation}
called Born effective charges, are gauge invariant~\cite{Resta94p899}.
Although the polarization $\vec{P}$ itself is not gauge invariant,
we can define the gauge invariant {\em change}~\cite{Rabe92p147} in polarization $\Delta \vec{P}$
along a path $C$ in the configuration space $\bi \Re$ as follows:
\begin{equation}
\Delta P_{\alpha} = \frac{1}{V}\sum_{i\beta}\int _C Z^*_{i,\alpha \beta}\mathrm{d}\Re _{i\beta}
\protect\label{eq:DELTAQ}
\end{equation}
The sufficient (but maybe not necessary)
conditions for $\Delta P_{\alpha}$ to be independent of the path
are given by Stokes' theorem. Let us consider two different paths connecting
two points in the configurational space. If there is at least one hyper-surface
bounded by the closed loop formed by these two paths, on which $Z^*_{i,\alpha \beta}$ are
differentiable at every point, then, according to Stokes' theorem,
the integral over the closed loop is zero, so $\Delta \vec{P}$ does not depend on the path. The Born effective
charges are not differentiable in insulator-metal transition regions of the 
configurational space. Therefore, $\Delta \vec{P}$ is the same along any two insulating
paths that can be continuously deformed to each other without crossing a metallic region.

It is interesting to note
that model systems can be constructed,
in which the above condition is not satisfied. In such systems, a closed
loop in a parameter space can result in electron transfer, leading to 
quantum adiabatic electron transport~\cite{Vanderbilt93p4442, Thouless83p6083,Niu86p5093,Niu90p1812} 
without net nuclear current. Since we are interested in identifying ions in solids here,
we limit ourselves
to considering atomic displacements whose length is larger than
the electron localization length, and the displacements of other atoms are all localized. This is 
always achievable by choosing a large enough unit cell to prevent interactions
between displaced sublattices.
 Two insulating paths in such ``dilute limit'' cannot form a loop that
leads to electron transfer, because electrons in an insulator are localized~\cite{Blount62p305}, and 
such electron transfer would mean that at some point on the path there would be delocalized electrons in insulating media.
This is not possible without crossing the band gap and causing a metallic state on the loop. Thus, for our purposes,
it is enough to find only one insulating path in parameter space. Indeed, while atoms could move in different environments
along two different paths, the vacancies left behind would stay in essentially 
the same environment, but would have different charges during some parts of the two
paths, if the charge transferred by the same atom along the two paths were different.
If  the system stays insulating along path 1 for which
the vacancy has larger number of electrons, it cannot stay insulating also along
path 2, because extra electrons should cross the band gap
at some point along path 2.

(ii) {\em $\Delta \vec{P}$ is parallel to $\vec{R}$, the lattice
vector by which the sublattice is displaced.} 
Choose a unit cell defined by $\vec{R}$ and two other lattice
vectors  $\vec{R}_2$ and $\vec{R}_3$.
A supercell for the same physical crystal can be 
defined by lengthening the unit cell along $\vec{R}_2$ by a factor
of $m$, but keeping
the same dimensions along $\vec{R}$ and $\vec{R}_3$. The  supercell
contains the original sublattices plus their images at $k\vec{R}^{\prime}_2/m$,
($k=1, 2, \ldots, m-1$), where
$\vec{R}^{\prime}_2$ = $m\vec{R}_2$ is the new lattice vector,
and the volume of the
supercell is increased to $V^{\prime}$ = $mV$. If we move
all of these sublattices (successively or together),
the situation is equivalent to the previous
one, and $\Delta \vec{P}$ is the same. But if we move one of these
sublattices by $\vec{R}$, the new polarization change
\begin{equation}
        \Delta \vec{P}^{\prime}=\frac{e}{V^{\prime}}\left (
        n^{\prime}\vec{R}+n^{\prime}_2\vec{R}^{\prime}_2
        +n^{\prime}_3\vec{R}_3 \right )
\end{equation}
must be $\Delta \vec{P}/m$ by symmetry. Thus,
\begin{eqnarray}
\Delta \vec{P}^{\prime}=\frac{e}{mV}\left (n^{\prime} \vec{R} + m n^{\prime}_2 \vec{R}_2 + n^{\prime}_3 \vec{R}_3 \right )=\nonumber \\
\frac{1}{m}\frac{e}{V} \left (n \vec{R} + n_2 \vec{R}_2 + n_3 \vec{R}_3 \right )=\frac{1}{m}\Delta \vec{P}
\end{eqnarray}
Since $\vec{R}$, $\vec{R}_2$ and $\vec{R}_3$ are linearly independent, it immediately follows
 that $n^{\prime}$ = $n$, $n^{\prime}_2$ = $n_2/m$, and
$n^{\prime}_3$ = $n_3$. 
Since $n_2$ must be divisible by every integer $m$, we get $n_2$ = 0.
Similarly, we can prove that $n_3$ = 0. 
Therefore displacement of a sublattice by $\vec{R}$ creates a polarization change
 $\Delta \vec{P}$ = $Ne\vec{R}/V$, directed along $\vec{R}$.

\begin{figure*}
\includegraphics[width=0.95\textwidth]{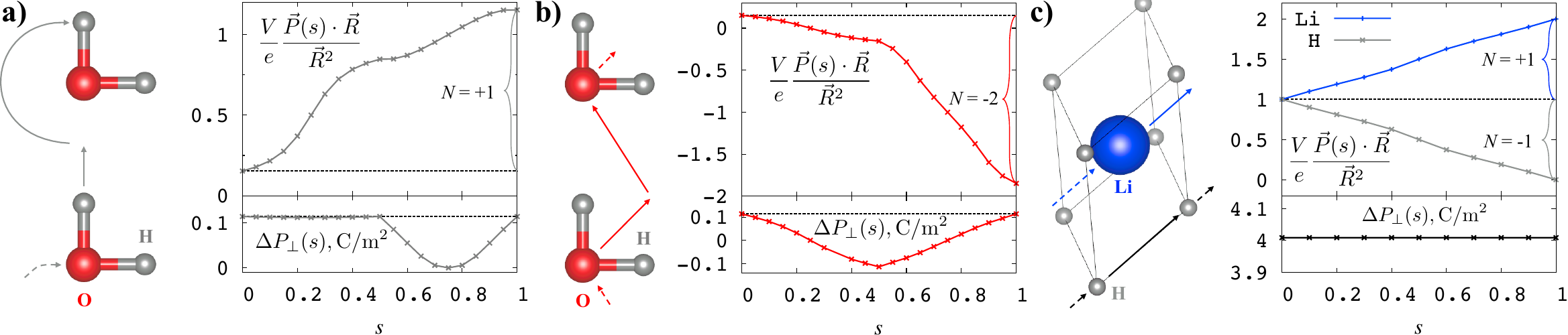}
\caption{Oxidation state $N$ for hydrogen (a) or oxygen (b)
in the ice model and
for Li or H in LiH (c). 
\protect\label{fig:H2O}}
\end{figure*}

(iii) {\em The quantity which we deem as  oxidation state}
\begin{equation}
N=\frac{V}{e}\frac{\Delta \vec{P}\cdot \vec{R}}{\vec{R}^2}
\label{eq:N}
\end{equation}
{\em is always an integer}. This is evident  from the conclusion of (ii): $\Delta \vec{P}$ = $Ne\vec{R}/V$.
Since polarization is proportional to the dipole
moment of the unit cell, $\Delta \vec{P}$ is the change in the dipole moment
upon transfer of an atom. This change is
directed along the vector connecting initial and final positions of the atom,
which can be interpreted as the change in dipole moment due to transfer
of a constant charge $Ne$. 
Note also that according to (i) $N$
is the same regardless of particular path of sublattice displacement.

Moreover,  in the Berry phase expression, polarization
can be written in terms of Wannier
function centers (WCs) ~\cite{Zak89p2747,Kingsmith93p1651}. 
This allows for mapping of wavefunctions to point charges and
restoring the classic ionic model.
The above conclusions then imply that transferring an atom by a lattice
vector results in transferring some of the WCs
by the same lattice vector, while the rest of the WCs
remain in the starting unit cell.  The Berry phase of the wavefunctions
carries the information on how many WCs move together with
any particular atomic sublattice. In this framework, the physical meaning of OS defined in
this method parallels the traditional definition as partitioning electrons (WCs) to atoms.

(iv) {\em The value of $N$ is the same, regardless of which lattice
vector $\vec{R}$ is chosen for sublattice displacement. $N$ in
equation~(\ref{eq:N})
depends only on the atomic species and its environment.} 
Assume a unit cell defined by $\vec{R}_1$, $\vec{R}_2$ and $\vec{R}_3$ exists,
for which $N$ along $\vec{R}_1$ and $\vec{R}_2$ are
different for a given sublattice. Consider sequential displacements
of the sublattice first along $\vec{R}_1$, and
then along $\vec{R}_2$. According to the proof above, the total polarization
change $\Delta \vec{P}$ should be directed along $\vec{R}_1+\vec{R}_2$.
However, since $\Delta \vec{P}$ is equal to the sum of
polarization changes $\Delta \vec{P}_1$ and $\Delta \vec{P}_2$ along
$\vec{R}_1$ and $\vec{R}_2$, the total polarization change
$\Delta \vec{P}$ = $\Delta \vec{P}_1$ + $\Delta \vec{P}_2=N_1e\vec{R}_1/V+N_2e\vec{R}_2/V$ 
cannot be parallel to $\vec{R}_1+\vec{R}_2$ if $N_1$ and $N_2$
are different, hence $N_1$ = $N_2$.

These four observations provide the basis for our definition of
 ions in solids. In the dilute limit, this formulation provides a
rigorous connection between polarization and oxidation state
associated with
a particular sublattice, as well as a tractable procedure for electron
partitioning among the ions in a solid. The partitioning is not based on
spatial proximity of WCs to a given nucleus, but rigorously derived
from the topology of electronic states. Note that the dilute limit provides a 
sufficient condition of the validity of (ii) to (iv), but may not be necessary.

To illustrate this new methodology, we calculate OS 
for atoms in ice, LiH, BaBiO$_3$ and Sr$_2$FeWO$_6$. 
We perform DFT calculations using the norm-conserving non-local pseudopotential
plane wave method. The electronic structure of the first two materials is calculated
with the generalized gradient approximation (GGA) to the exchange-correlation functional,
 as implemented in the \textsc{Abinit} package ~\cite{Gonze09p2582, Gonze05p558}.
The local density approximation (LDA) with the Hubbard $U$ parameter, as implemented
in the \textsc{Quantum-Espresso} package~\cite{Giannozzi09p395502},
 is used to calculate the electronic structure of BaTiO$_3$ and Sr$_2$FeWO$_6$.

Water molecules in ice are rearranged for simplicity. Namely, 
each 4~\AA{} $\times$ 4~\AA{} $\times$ 3~\AA{} tetragonal unit cell
contains one water with H-O-H of 90$^\circ$ and O-H 
bond lengths of 1~\AA. The oxygen atom is placed at the origin, and the
hydrogen atoms are on $\hat{y}$ and $\hat{z}$ axes.

We transfer the hydrogen atom on the $\hat{z}$ axis to the next unit cell
along $\hat{z}$. The path (Fig.~\ref{fig:H2O}a) consists of  
a 1~\AA{} straight line and a 1~\AA{} radius half-circle in the H-O-H plane.
$N$ continuously increases from zero to +1, meaning it is H$^+$,
while the overall change in the perpendicular component of $\vec{P}$ is zero.

Fig.~\ref{fig:H2O}b shows $N$ = -2 for the oxygen
atom. In this case, the oxygen sublattice is first moved straight
along the H-O-H bisector for $\sqrt{2}$~\AA, during which hydrogen 
atoms also move in the H-O-H plane to keep the O-H bond lengths unchanged. Next, the hydrogen
 atoms return to their original positions,
and the oxygen atom is moved straight to the starting O location in the next unit cell.

In order to demonstrate that our formulation of oxidation state
is a function of crystal environment,
we perform calculations for LiH, whose
conventional OS of hydrogen 
is -1. The
results of calculations are shown in Fig.~\ref{fig:H2O}c.
LiH has the rock-salt crystal structure, with two atoms 
in the unit cell, and FCC lattice vectors. 
For each atom, the path corresponds to the transfer 
along one of the lattice vectors. It can be easily seen from the
figure that $N=-1$ for hydrogen, revealing H$^-$, while for Li $N=+1$,
signifying Li$^+$.

\begin{figure}
\includegraphics[width=0.45\textwidth]{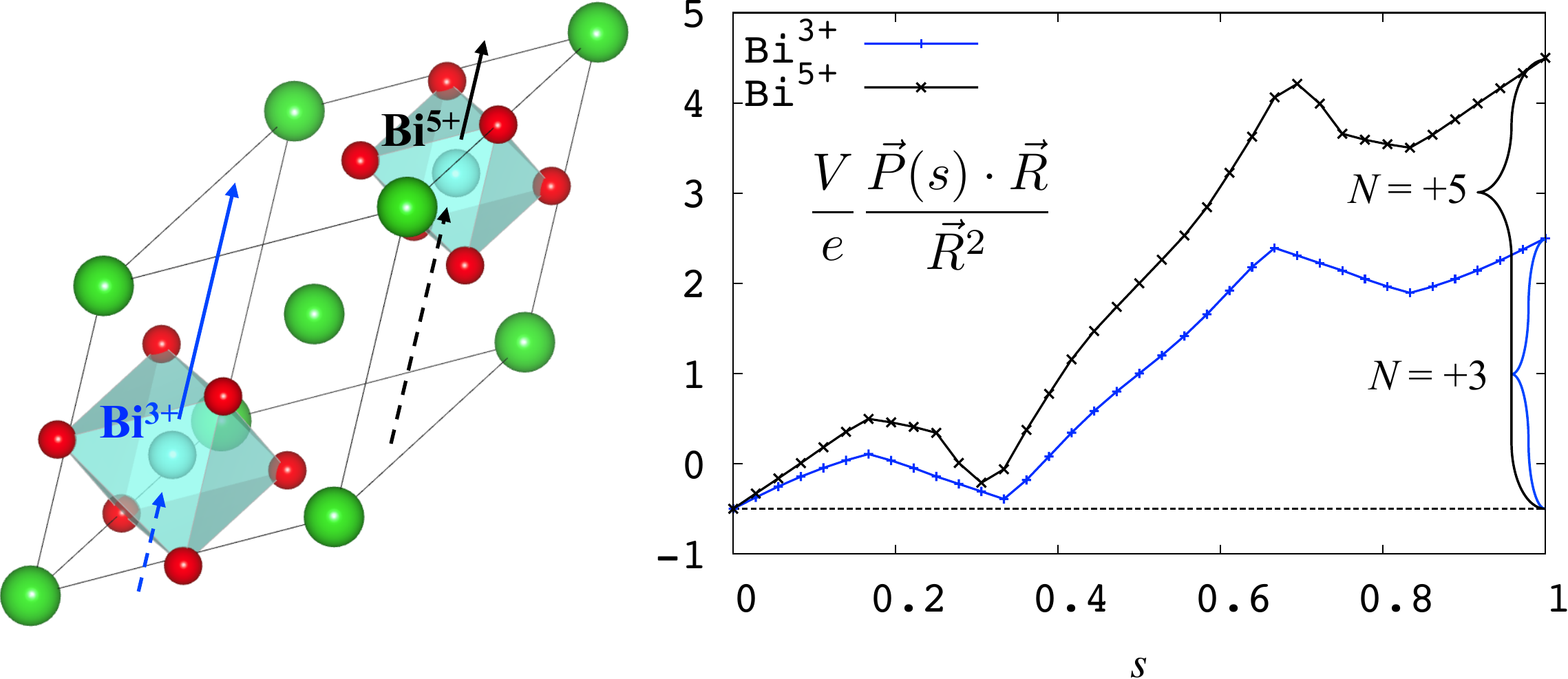}
\caption{Oxidation state $N$ for Bi$^{5+}$ (black) or Bi$^{3+}$ (blue) in 
Ba$_2$Bi$^{5+}$Bi$^{3+}$O$_6$.
\protect\label{fig:BBO}}
\end{figure}

Next,  the BaBiO$_3$ system is chosen to demonstrate our method's ability to differentiate
Bi$^{5+}$ and Bi$^{3+}$ in a single phase. The formal OS of Bi for cubic perovskite
BaBiO$_3$ is $+4$, leading to a charge disproportionation $2$Bi$^{4+}$ $\rightarrow$
Bi$^{5+}$ $+$ Bi$^{3+}$ which is coupled to a collective oxygen
octahedral breathing mode\cite{Thonhauser06p212106}. Therefore, a double perovskite 10-atom rhombhedral
unit cell is used with conventional lattice parameter of 8.66~\AA{} and octahedral breathing giving
 Bi-O bond lengths of 1.97~\AA{} and 2.37~\AA. 
 An effective Hubbard $U$ term of 6~eV is applied to the oxygen $2p$
orbitals to correct for the band gap underestimation of DFT; this
results in an LDA+$U$ band gap of 1.88 eV, much larger than the
published GGA-PBE band gap of 0.6 eV and close to the measured optical
band gap of 2.05~eV\cite{Tang07p12779}.

Insulating paths for each Bi sublattice are found by moving one Bi cation and its images
straight along a lattice vector and carefully adjusting positions of surrounding O ions to
minimize Bi-O bond length change during the path. The results in Fig.~\ref{fig:BBO} show that Bi in the smaller
oxygen cage has $N$ = +5 while the other Bi has $N$ = +3, consistent with conventional wisdom
that Bi$^{5+}$ has a smaller ionic radius than Bi$^{3+}$.

\begin{figure}
\includegraphics[width=0.45\textwidth]{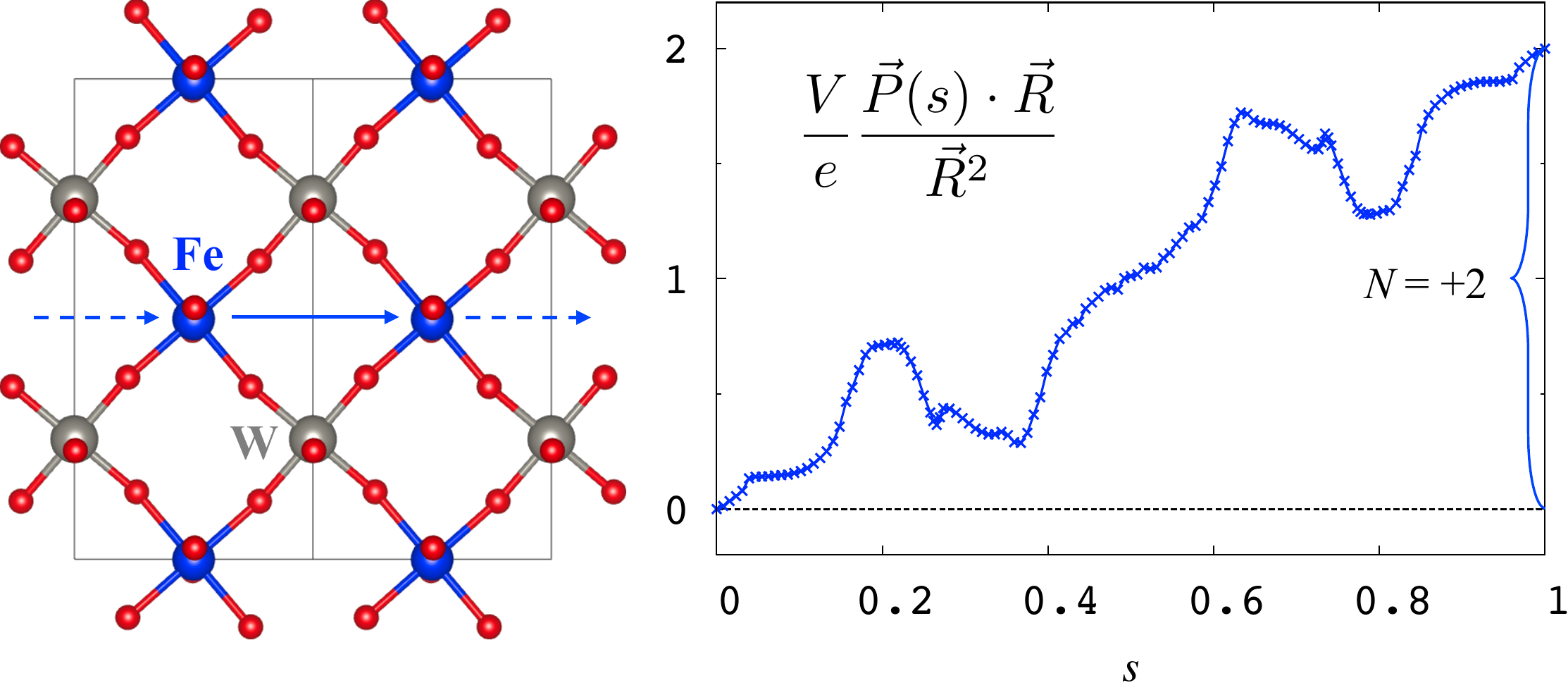}
\caption{Oxidation state $N$ for Fe$^{2+}$ in 
Sr$_2$Fe$^{2+}$W$^{6+}$O$_6$.  One FeWO$_4$ layer and the sublattice displacement vector
are shown.
\protect\label{fig:SFW}}
\end{figure}

Lastly, we use Sr$_2$FeWO$_6$ to show that for a system with variable OS elements (Fe$^{2+}$ vs. Fe$^{3+}$),
this method can definitively assign OS and resolve the ambiguity. While most of Sr$_2$Fe\textit{M}O$_6$ (\textit{M} = Ta, Mo, Re, \ldots)
are ferromagnetic metals and contain Fe$^{3+}$, Sr$_2$FeWO$_6$ is an anti-ferromagnetic insulator~\cite{Kawanaka99p2890}. Fe charge
state in Sr$_2$FeWO$_6$ has been studied experimentally by M\"{o}ssbauer spectroscopy and the  
results fall borderline between high spin $+2$  and low spin $+3$ state. Base on the large unit cell and a  purposed ``cancellation effect''
it is assigned to be
 Fe$^{2+}$~\cite{Kawanaka00p581}. In our calculations, we used the experimental 80-atom structure of monoclinic Sr$_2$FeWO$_6$
  unit cell, which accommodates the 
 type-$\mathrm{II}$ anti-ferromagnetic ordering of Fe~\cite{Matteo03p184427}. An effective Hubbard $U$ of 4~eV is added to the strongly correlated Fe $3d$ orbitals~\cite{Fang01p180407}. An Fe ion is moved along the shortest
axis, and the path is initially obtained by the nudged elastic band (NEB) method, with a subsequent manual adjustment of oxygen positions
 for bond length preservation,
and addition of more intermediate structures to ensure continuity. 
The result in Fig.~\ref{fig:SFW} indicates that Fe has $N$ = +2, as earlier work stated, and
shows the ability of our methodology to identify OS for systems where ambiguity arises from multiple variable-valent elements.

\begin{figure}[!tbh]
\includegraphics[width=0.45\textwidth]{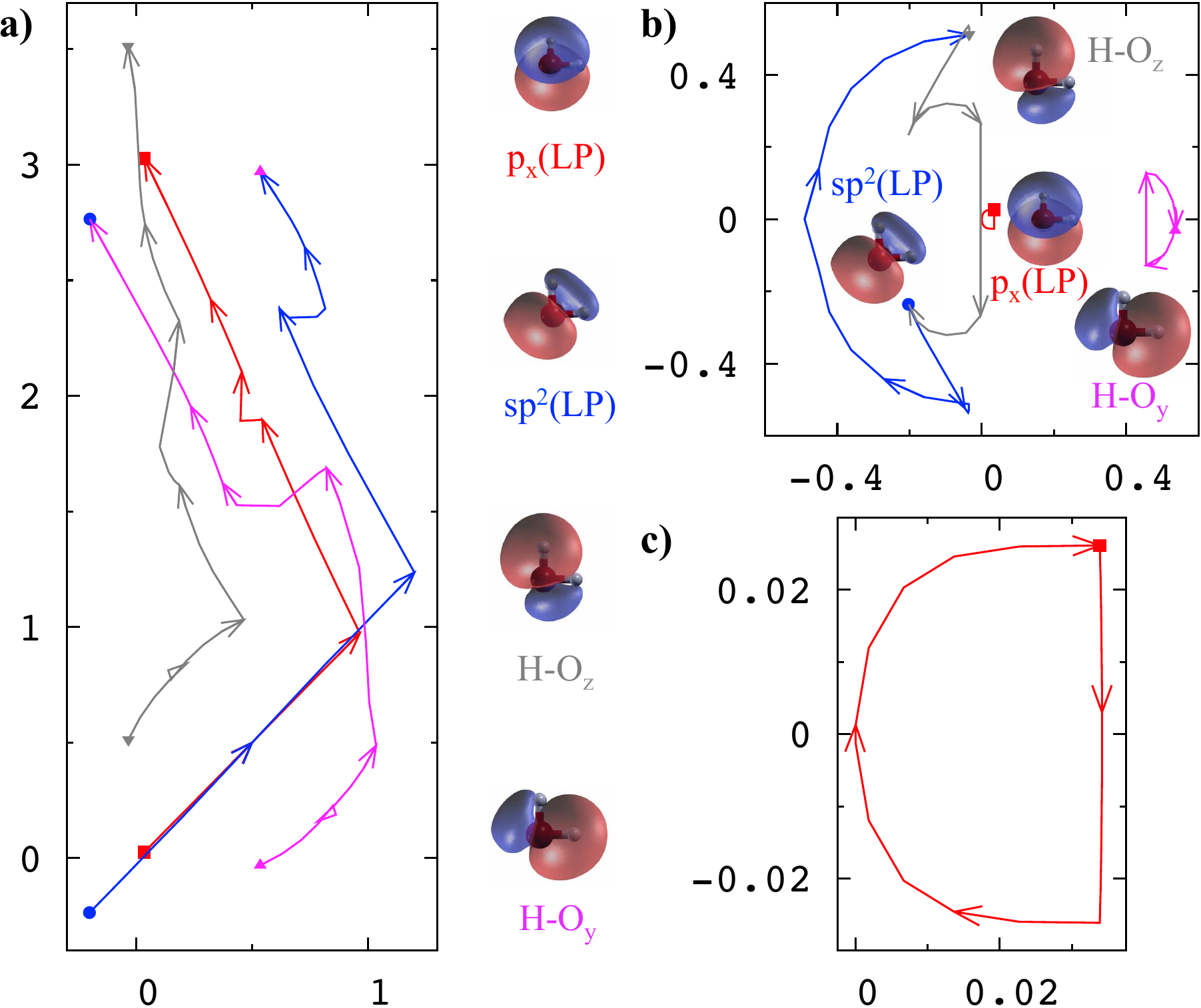}
\caption{Trajectories of the maximally localized WCs
generated by the motion of the O (a) and H 
(b) in the H$_2$O system. The O 2$p_x$ WC trajectory from (b)
is shown on an expanded scale in (c). The units are \AA.  
\protect\label{fig:H2Owanc}}
\end{figure}

 To visualize the idea of ``electrons topologically bound to nucleus'', we calculate the trajectories of the centers of maximally localized 
Wannier functions~\cite{Marzari97p12847, Boys60p296} for the H$_2$O system with 
the Wannier90 program~\cite{Mostofi08p685}.
Fig.~\ref{fig:H2Owanc}a
shows the results when the O atom is moved along
the aforementioned path. 
Note that along the path, high symmetry structures are intentionally avoided due to the 
discontinuity that would otherwise occur in the trajectories. As can be seen, all WCs move with 
the O atom to the next unit cell. The WC
which we assign as the oxygen 2$p_x$ orbital stays very close to the
nucleus along the whole path, so that its trajectory almost coincides
with the trajectory of the nucleus. 
WCs can also exchange their characters
along the way depending on the path: the O $sp^2$ lone pair becomes O-H$_y$ bond, and vice versa. 
 When an H atom is moved,
however, all WCs stay in the same unit cell
(Fig.~\ref{fig:H2Owanc}b), so that there is no overall electron displacement.

In summary, we developed a method for partitioning electrons to atoms based on their wavefunction topologies.
When an atom moves to its image position in periodic insulators, change of polarization reveals the number of Wannier
function centers that move with the atom, provided that the system stays insulating. This effectively indicates the number
of electrons that ``belong'' to the nucleus and establishes a rigorous definition of oxidation state of ions in solids. This
concept of ``ions-in-solids'' can have important implications for materials modeling. For example, electron
redistribution upon defect formation or ion transport
through a polar medium can be described in terms
of ion deformation (no change in OS) or charge transfer (when OS changes).

\begin{acknowledgments}
L. J. was supported by the US AFOSR under grant FA9550-10-1-0248,
S. V. L. was supported by the US
DoE BES under grant DE-FG02-07ER15920,
and A. M. R. was supported by 
the US ONR under grant number
N00014-11-1-0578. Computational support was provided by
the US DoD. We thank Eugene J. Mele and Barri J.
Gold for fruitful discussions.
\end{acknowledgments}

\bibliography{rappecites}

\end{document}